\documentclass[pra,twocolumn,superscriptaddress,showpacs]{revtex4}
\usepackage{amssymb}\usepackage{graphicx}\usepackage{color}\usepackage{psfrag}
\begin{document}
\title{Frequency Bin Entangled Photons}
\author{L. Olislager}
\email{lolislag@ulb.ac.be}
\affiliation{Service OPERA-Photonique, CP 194/5, Universit\'e Libre de Bruxelles, Avenue F.D. Roosevelt 50, 1050 Brussels, Belgium}
\author{J. Cussey}
\affiliation{SmartQuantum SA, Espace Ph\oe nix, Route du Rad\^{o}me, 22560 Pleumeur-Bodou, France}
\author{A.T. Nguyen}
\affiliation{Service de Physique et Electricit\'e, Institut Meurice, Haute Ecole Lucia de Brouck\`ere, Avenue Emile Gryzon 1, 1070 Brussels, Belgium}
\author{Ph. Emplit}
\affiliation{Service OPERA-Photonique, CP 194/5, Universit\'e Libre de Bruxelles, Avenue F.D. Roosevelt 50, 1050 Brussels, Belgium}
\author{S. Massar}
\affiliation{Laboratoire d'Information Quantique, CP 225, Universit\'e Libre de Bruxelles, Boulevard du Triomphe, 1050 Brussels, Belgium}
\author{J.-M. Merolla}
\affiliation{D\'epartement d'Optique P.M. Duffieux, Institut FEMTO-ST, Centre National de la Recherche Scientifique, UMR 6174, Universit\'e de Franche-Comt\'e, 25030 Besan\c{c}on, France}
\author{K. Phan~Huy}
\affiliation{D\'epartement d'Optique P.M. Duffieux, Institut FEMTO-ST, Centre National de la Recherche Scientifique, UMR 6174, Universit\'e de Franche-Comt\'e, 25030 Besan\c{c}on, France}
\date{\today}
\begin{abstract}
A monochromatic laser pumping a parametric down conversion crystal generates frequency entangled photon pairs. We study this experimentally by addressing such frequency entangled photons at telecommunication wavelengths (around $1550\,\mathrm{nm}$) with fiber optics components such as electro-optic phase modulators and narrow band frequency filters. The theory underlying our approach is developed by introducing the notion of {\em frequency bin} entanglement. Our results show that the phase modulators address coherently up to eleven frequency bins, leading to an interference pattern which can violate a Bell inequality adapted to our setup by more than five standard deviations.
\end{abstract}
\pacs{42.50.Dv, 03.67.Bg, 03.65.Ud}
\maketitle

\section{Introduction}

Entanglement is one of the most fascinating aspects of quantum mechanics, used both for fundamental tests of physical principles and for applications such as Quantum Key Distribution (QKD). Many different kinds of photonic entanglement have been produced, including entanglement in polarization \cite{A1981, K1995}, momentum \cite{R1990}, angular momentum \cite{M2001} and time-energy. Investigation of the latter degree of freedom has been mainly inspired by two photon bunching experiments first carried out by Ou and Mandel \cite{O1988}, see \cite{T1990, L2009}; and by Franson's proposal \cite{F1989} for addressing the entanglement in the time domain, see \cite{B1992, K1993, T1998, B1999, T2000, T2002}. Photons entangled simultaneously in both time-energy and other degrees of freedom have also been studied \cite{S1994, B2005}. Time-energy entanglement can also be viewed as frequency entanglement, as demonstrated in recent works \cite{F2009, R2009}. Here we show how to address time-energy entangled photons directly in the frequency domain. This is realized in optical fibers by using commercially available telecommunication components.

Before presenting our approach, it may be useful to recall Franson's proposal \cite{F1989} which is based on three key ideas. First, a continuous pump laser produces time entangled photon pairs: the emission time of each photon is uncertain, but both photons are emitted simultaneously. Second, one uses measurements that resolve the arrival time of the photons. This leads to the concept of {\em time bin}: two photons whose arrival time cannot be distinguished by the detectors belong to the same time bin. Third, different time bins are made to interfere by using unbalanced Mach-Zehnder interferometers. These ideas provide a powerful platform to investigate quantum entanglement, yielding seminal works such as long distance violation of Bell inequalities \cite{T1998} and entanglement based QKD \cite{T2000}.

Here we also use the time-energy degree of freedom, but the way it is addressed is very different. However at the conceptual level there is an instructive parallel between our approach and that of Franson. First, a narrow band pump laser produces frequency entangled photon pairs: the frequency of each photon is uncertain, but the sum of the frequencies is well defined. Second, our detectors are preceded by narrow band filters that resolve the frequency of the detected photons. This leads to the concept of {\em frequency bin}: two photons whose frequency is so close that they
cannot be distinguished by the filters are said to lie in the same frequency bin. Third, different frequency bins are made to interfere by using electro-optic phase modulators.

Our work is inspired by, or related to, earlier proposals for manipulating qubits in the frequency domain \cite{M9901, M9902, B2007, C2008, S2003, H2004, H2005}. Our experimental techniques follow closely those of QKD systems in which the quantum information is encoded in frequency sidebands of an attenuated coherent state \cite{M9901, M9902, B2007}. Such systems allow efficient transmission of quantum information at telecommunication wavelengths. The main advantage of this method for encoding and carrying out transformations on optical qubits is that one does not need to stabilize paths in optical interferometers. Rather one must only lock the local Radio Frequency (RF) oscillators used by Alice and Bob, which is much easier. Furthermore information encoded in sidebands is unaffected by birefringence in the optical fiber used for transmission. Recent improvements to these experiments have included dispersion compensation and long distance synchronization of the sender and receiver \cite{C2008}, so that this approach constitutes now the only commercial alternative to time bin based QKD.

The architecture reported in \cite{M9901, M9902} was dedicated to QKD using faint laser pulses, but it is inefficient when single photons are used because weak modulation amplitudes are required. To overcome this limitation, an alternative method was proposed \cite{B2007} in which information is encoded both in the amplitude and relative phase of three frequency bands generated by electro-optic phase modulators. This second approach is attractive because in principle the phase modulators need not attenuate the signal, because there is no need for a strong reference pulse, and because the phase modulators can address many frequency sidebands simultaneously. Here we transpose, with appropriate modifications, the setup of \cite{B2007} to the entangled photon case.

In the following we first describe our experiment and give the principle of our method. Then we present our experimental results, and demonstrate that the phase modulators can coherently address up to eleven frequency bins. Note that high dimensional entanglement has been studied in a number of earlier experiments, see e.g. \cite{S1994,B2005,V2002,T2004,R2004,O2005,E2007}. We finally show that the two photon interference pattern we obtain can in principle violate a Bell inequality adapted to our setup by more than five standard deviations.

\section{Experimental Setup, Theoretical Description, and Results}

\begin{figure}[ht]
\includegraphics[scale=.2]{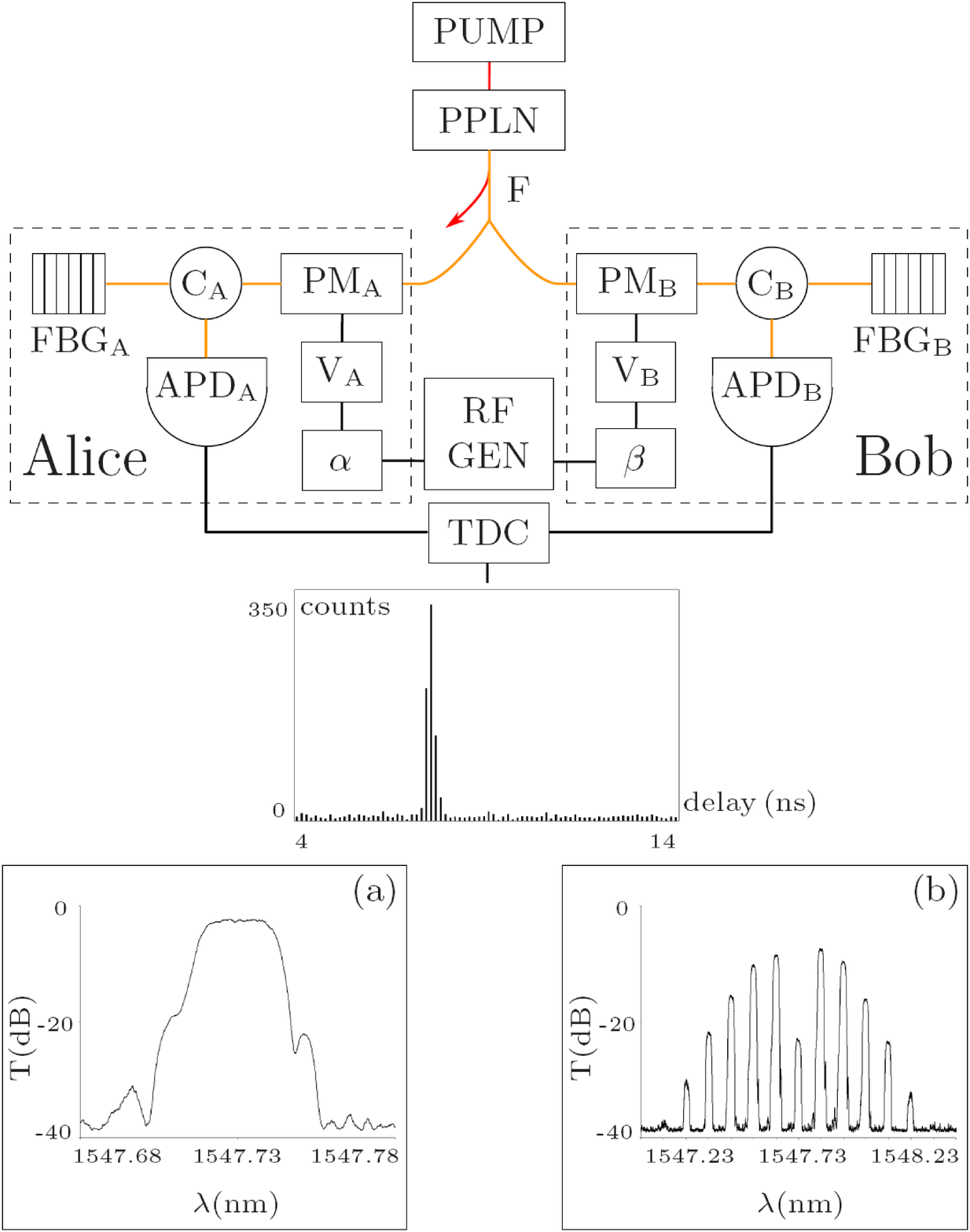}
\caption{\label{fig1}Experimental setup. The quasi-monochromatic pump laser (PUMP) creates photon pairs in the Periodically Poled Lithium Niobate waveguide (PPLN) and is then removed by a drop Filter (F). Alice and Bob's photons are selected passively by a $3\,\mathrm{dB}$ splitter. The photons then pass through electro-optic Phase Modulators ($\mathrm{PM_{A,B}}$). The phase modulators are driven by a $12.5\,\mathrm{GHz}$ Radio Frequency Generator (RF GEN) whose output is controlled by variable attenuators ($\mathrm{V_{A,B}}$) and phase shifters ($\alpha,\,\beta$). RF isolators (not shown) guarantee the independence of Alice and Bob's settings ($75\,\mathrm{dB}$ isolation). Individual frequency bins are selected by narrow band filters which consist of a Fiber Bragg Grating ($\mathrm{FBG_{A,B}}$) preceded by a Circulator ($\mathrm{C_{A,B}}$). The photons are finally detected by Avalanche Photo-Diodes ($\mathrm{APD_{A,B}}$) and the electronic signals sent to a Time to Digital Converter (TDC) to perform a coincidence measurement. A typical figure recorded by the TDC, consisting of a background due to accidental coincidences and a narrow peak when the photons arrive in coincidence, is shown. Typical values for the Signal to Noise Ratio (SNR) (number of non accidental coincidences divided by number of accidental coincidences) are $\mathrm{SNR}\approx 100$. Inset (a) shows a typical transmission spectrum of the narrow band filters. Inset (b) illustrates how the phase modulators generate new frequencies: if a broadband source passes through the narrow band filter (as in (a)), and then through a phase modulator, one obtains the spectrum of (b). The height of the peaks is given by the norm square of the coefficients in Eq. (\ref{eq1}). When $a\approx 2.74$ (corresponding to the maximum RF power produced by the source) one can see eleven frequency bins corresponding to the order of the peak $p$, see Eq. (\ref{eq1}), ranging from $p=-5$ to $p=+5$. In insets (a) and (b), $\lambda$ is the wavelength and T the transmission. Note that the horizontal scale is different in insets (a) and (b).}
\end{figure}

Our all fiber experiment is schematized in Fig. \ref{fig1}. Photon pairs are generated by parametric down conversion in a $3\,\mathrm{cm}$ long Periodically Poled Lithium Niobate (PPLN) waveguide (HC Photonics). PPLN waveguides have emerged as the preferred photon pair source at telecommunication wavelengths because of their extremely high spectral brightness \cite{T2001, H2008}. The narrow band pump laser (Sacher Lasertechnik, $\lambda_p=773.865\,\mathrm{nm}$, $P_p\approx6\,\mathrm{mW}$) is removed with a drop filter F insuring $125\,\mathrm{dB}$ isolation. The identically polarized photon pairs, distributed around $\lambda_0=1547.73\,\mathrm{nm}=2\pi\mathrm{c}/\omega_0$, are separated with a $3\,\mathrm{dB}$ coupler. Interesting cases occur when the photon pair is split: one photon is sent to Alice (A) and the other to Bob (B). At the output of the coupler, the photons pass through electro-optic Phase Modulators $\mathrm{PM_{A,B}}$ (EOSPACE, $25\,\mathrm{GHz}$ bandwidth, $2.5\,\mathrm{dB}$ loss, half-wave voltage $V_\pi\approx2.9\,\mathrm{V}$), whose active axis are aligned with the linear polarization of the photons (preserved thanks to polarization maintaining fiber components), and to which are applied sinusoidally varying voltages at frequency $\Omega/2\pi=12.5\,\mathrm{GHz}$, with amplitudes $\mathrm{V_{A,B}}$ and phases $\alpha,\,\beta$ which can be controlled. The induced time dependent optical phases $\phi_A(t)=a\cos(\Omega t-\alpha)$ and $\phi_B(t)=b\cos(\Omega t-\beta)$, where $a=\pi V_A/V_\pi$ and $b=\pi V_B/V_\pi$, lead to the unitary transformations
\begin{eqnarray}
\label{eq1}|\omega\rangle_A&\rightarrow&\sum_{p \in \mathbb{Z}}^{}|\omega+p\Omega\rangle_A U_p(a,\alpha)\,,\\
\label{eq2}|\omega\rangle_B&\rightarrow&\sum_{q \in \mathbb{Z}}^{}|\omega+q\Omega\rangle_B U_q(b,\beta)\,,
\end{eqnarray}
where subscripts A and B refer to Alice and Bob's photons, $U_p(a,\alpha)=J_p(a)e^{ip(\alpha-\pi/2)}$, $U_q(b,\beta)=J_q(b)e^{iq(\beta-\pi/2)}$, and $J_{p,q}$ is the $p,q\,$th-order Bessel function of the first kind. The range of values of $a,b$ which were experimentally accessible are limited to $\{0,2.74\}$ due to the finite power of the RF generator used. The photons are then sent through narrow band filters $\mathrm{F_{A,B}}$ which consist of a Fiber Bragg Grating (FBG) preceded by a circulator. Losses are $0.2\,\mathrm{dB}$ for the FBGs and $0.8\,\mathrm{dB}$ for the circulators (round-trip). The spectral characteristics of the FBGs are: Full Width at Half Maximum (FWHM) $\approx3\,\mathrm{GHz}$, and more than $30\,\mathrm{dB}$ isolation at $6.25\,\mathrm{GHz}$, see Fig. \ref{fig1} inset (a). Alice's filter is kept fixed on angular frequency $\omega_A=\omega_0$. It is athermally packaged to reduce central wavelength deviation to $1\,\mathrm{pm}/\mathrm{K}$. The temperature of Bob's filter is controlled by a Peltier module, which allows continuous tuning of the reflected frequency $\omega_B$ over a $1\,\mathrm{nm}$ range around $\omega_0$. The use of such narrow band filters together with a spectrally bright PPLN source of entangled photons has been reported previously in the context of four-photon experiments \cite{H2007, H2008}. Finally the photons are detected by two Avalanche Photo-Diodes $\mathrm{APD_{A,B}}$ operated in gated mode (id Quantique, efficiency $15\%$, dark count rates $3.5\cdot10^{-5}/\mathrm{ns}$ and $8.0\cdot10^{-5}/\mathrm{ns}$) and a time to digital converter performs a coincidence measurement. The maximum coincidence rate was approximately $10\,\mathrm{Hz}$, which is consistent with the earlier work of \cite{H2007, H2008}.

A parametric down conversion source pumped by a monochromatic beam produces an entangled state which we can idealize as
\begin{equation}\label{eq3}
|\Psi\rangle=\int\mathrm{d}\omega|\omega_0+\omega\rangle_\mathrm{A}|\omega_0-\omega\rangle_\mathrm{B}\,.
\end{equation}
The total energy of the photon pair is well defined, but the energy of each photon is uncertain. For simplicity of notation we have not normalized Eq. (\ref{eq3}). This does not affect our predictions as we are interested in the {\em ratios} of the probabilities of finding photon A at one frequency and photon B at another frequency for different settings $a,b,\alpha,\beta$ of the phase modulators. For a discussion of how to normalize Eq. (\ref{eq3}) so as to describe a rate of photon pair production, see the appendix.

Note that taking the Fourier transform of Eq. (\ref{eq3}) would yield a description of the state in terms of time entanglement: the arrival time of each photon is uncertain, but the difference between the arrival time of Alice and Bob's photon is well defined. The approximations leading to Eq. (\ref{eq3}) consist in neglecting the finite pump bandwidth (which is approximately $2\,\mathrm{MHz}$) and the finite signal and idler photons bandwidths (which are approximately $5\,\mathrm{THz}$). This is legitimate as they are respectively much smaller than the bandwidths of the filters $\mathrm{F_{A,B}}$ (which are approximately $3\,\mathrm{GHz}$, see Fig. \ref{fig1} inset (a)) and much larger than the bandwidth sampled by the phase modulators (which is approximately $125\,\mathrm{GHz}$, see Fig. \ref{fig1} inset (b)).

According to Eqs (\ref{eq1},\ref{eq2}), the phase modulators realize interferences between photons whose frequencies are separated by integer multiples of $\Omega$. They thus play the same conceptual role as the Mach-Zehnder interferometers in Franson's scheme which realize interferences between different time bins. Using Eqs (\ref{eq1},\ref{eq2}) one can readily compute how the entangled state Eq. (\ref{eq3}) is affected by the phase modulators:
\begin{equation}\label{eq4}
|\Psi\rangle\rightarrow\int\mathrm{d}\omega'\sum_{d\in \mathbb{Z}} |\omega_0+\omega'\rangle_\mathrm{A}|\omega_0-\omega'+d\Omega\rangle_\mathrm{B}~c_{d}(a,b,\alpha,\beta)\,,
\end{equation}
with $\omega'=\omega+p\Omega$, $d=p+q$, and
\begin{equation}\label{eq5}
c_d(a,b,\alpha,\beta)=\sum_p^{}U_p(a,\alpha)U_{d-p}(b,\beta)\,.
\end{equation}
According to Eq. (\ref{eq4}) we will observe coincidences between Alice and Bob's photons only if the frequency bins $\omega_{A,B}$ in which they are detected are separated by integer multiples of $\Omega$: $\omega_A+\omega_B-2\omega_0=d\Omega$, $d\in \mathbb{Z}$.

The rate at which Alice and Bob will detect photons at angular frequencies $\omega_A=\omega_0+\omega'$ and $\omega_B=\omega_0-\omega'+d\Omega$ is proportional to
\begin{eqnarray}\label{eq6}
Q(\omega_0+\omega',\omega_0-\omega'+d\Omega|a,b,\alpha,\beta)=|c_d(a,b,\alpha,\beta)|^2\,.
\end{eqnarray}
(For a derivation of the proportionality factor, see the appendix.)

Because of the symmetries of Eqs (\ref{eq1},\ref{eq2},\ref{eq3}) the quantity Q depends only on the absolute value of the index $d$ (but not on the sign of $d$, nor on $\omega_0$ and $\omega'$) and on the phase difference $\Delta=\alpha-\beta$ (but not on $\alpha+\beta$):
\begin{eqnarray}\label{eq7}
Q(\omega_0+\omega',\omega_0-\omega'+d\Omega|a,b,\alpha,\beta)&=&Q(d|a,b,\Delta)\\
&=&Q(-d|a,b,\Delta)\,.\nonumber
\end{eqnarray}
The quantities $Q(d|a,b,\Delta)$ obey the normalization condition
\begin{equation}\label{eq8}
\sum_d Q(d|a,b,\Delta)=\sum_d |c_d(a,b,\alpha,\beta)|^2=1
\end{equation}
and satisfy
\begin{equation}\label{eq9}
Q(d=0|a,a,\pi)=1\quad,\quad Q(d\neq 0|a,a,\pi)=0\,.
\end{equation}
When the phase modulators are turned off ($a=b=0$), the correlations are trivial, and we have
\begin{equation}\label{eq10}
Q(d=0|a=b=0)=1\quad,\quad Q(d\neq 0|a=b=0)=0\,.
\end{equation}

\begin{figure}[ht]
\psfrag{d=0}{$d=0$}\psfrag{d=1}{$d=1$}\psfrag{d=2}{$d=2$}\psfrag{d=3}{$d=3$}\psfrag{d=4}{$d=4$}\psfrag{d=5}{$d=5$}
\psfrag{x0}[tr][cr][1][0]{\footnotesize{0}}
\psfrag{x1}[tr][cr][1][0]{\footnotesize{0.5}}
\psfrag{x2}[tr][cr][1][0]{\footnotesize{1}}
\psfrag{x3}[tr][cr][1][0]{\footnotesize{1.5}}
\psfrag{x4}[tr][cr][1][0]{\footnotesize{2}}
\psfrag{x5}[tr][cr][1][0]{\footnotesize{2.5}}
\psfrag{x6}[tr][cr][1][0]{\footnotesize{3}}
\psfrag{y0}[cr][cr][1][0]{\footnotesize{0}}
\psfrag{y1}[cr][cr][1][0]{\footnotesize{0.2}}
\psfrag{y2}[cr][cr][1][0]{\footnotesize{0.4}}
\psfrag{y3}[cr][cr][1][0]{\footnotesize{0.6}}
\psfrag{y4}[cr][cr][1][0]{\footnotesize{0.8}}
\psfrag{y5}[cr][cr][1][0]{\footnotesize{1}}
\includegraphics[scale=.2,trim=0 0 0 000mm,clip=true]{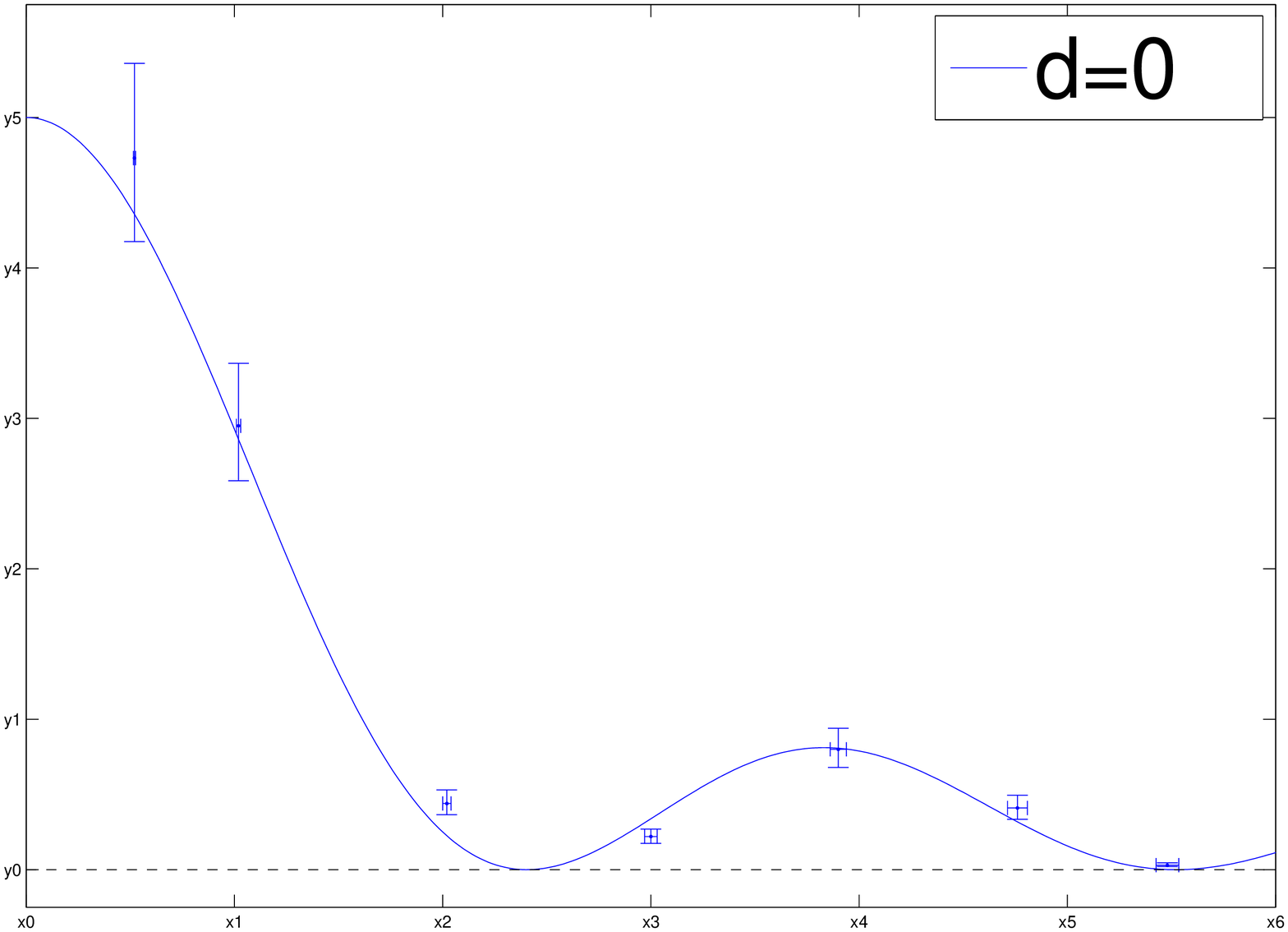}\\
\includegraphics[scale=.2,trim=0 0 0 120mm,clip=true]{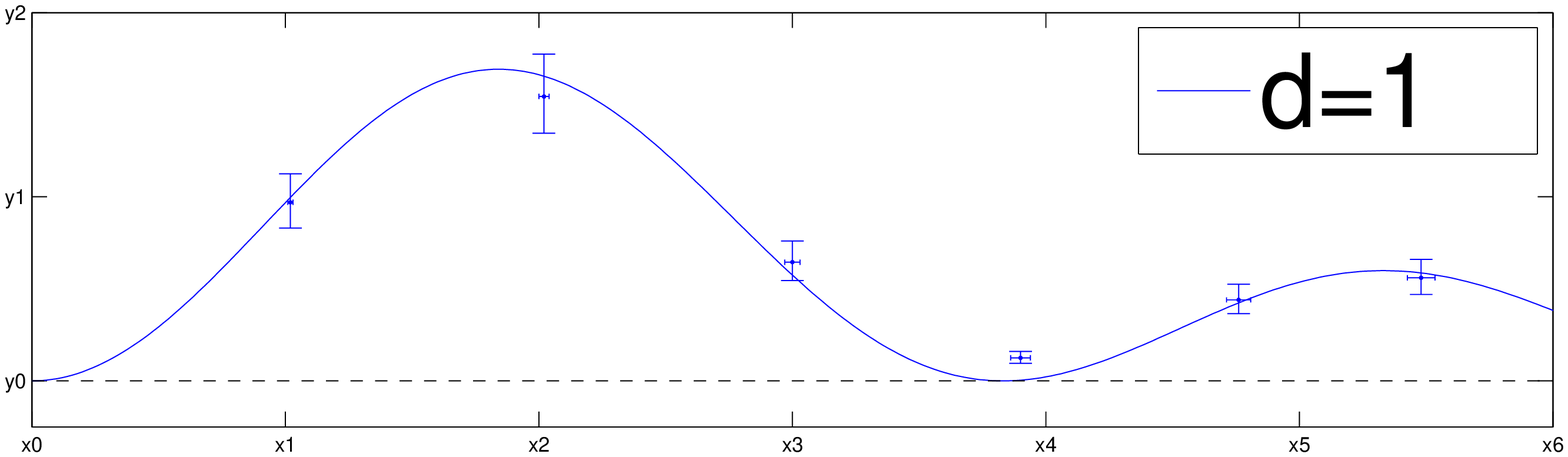}\\
\includegraphics[scale=.2,trim=0 0 0 120mm,clip=true]{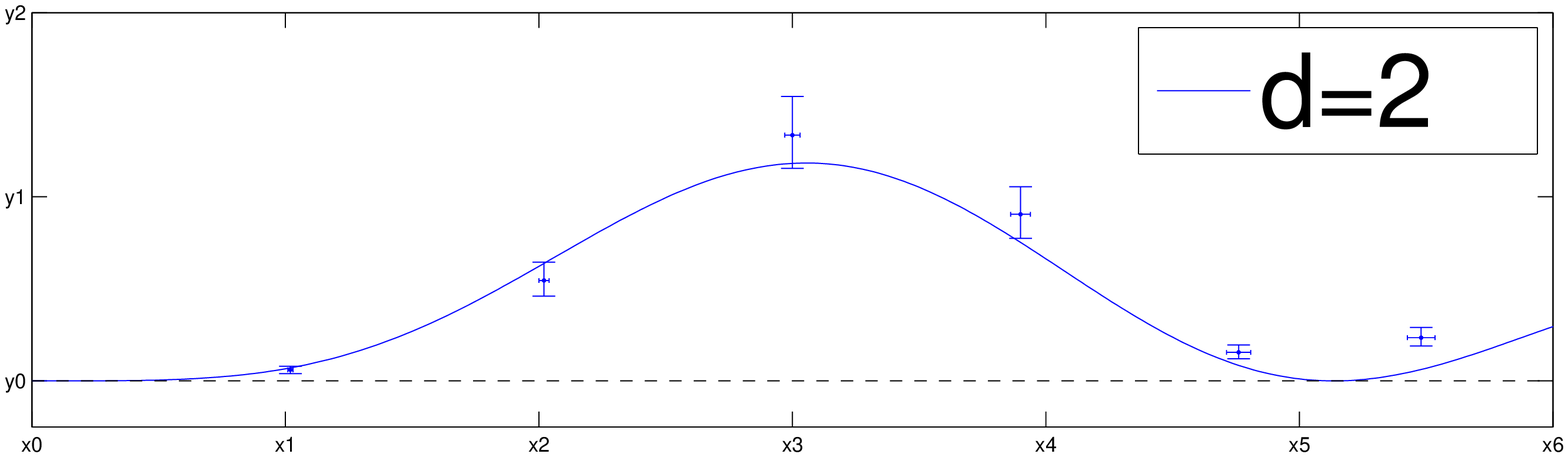}\\
\includegraphics[scale=.2,trim=0 0 0 138mm,clip=true]{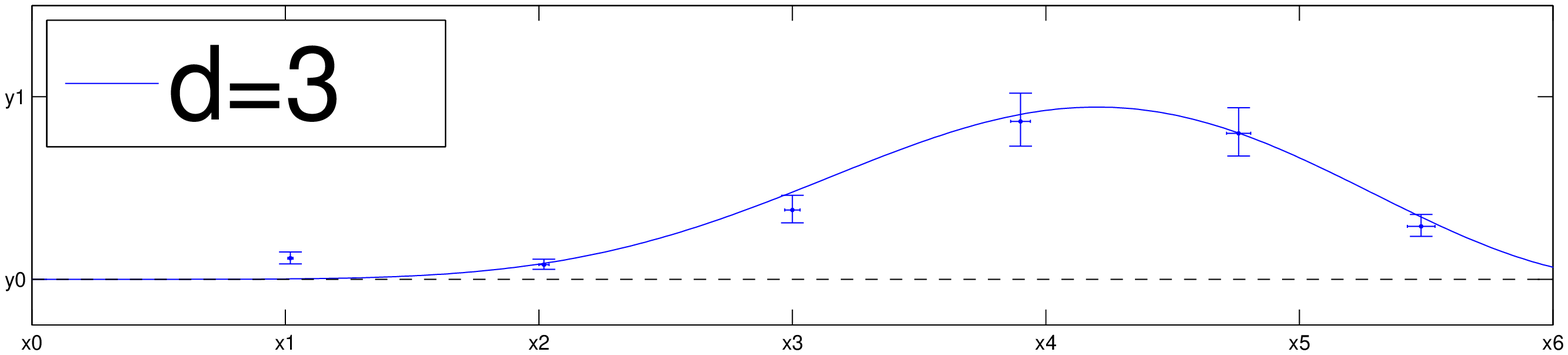}\\
\includegraphics[scale=.2,trim=0 0 0 156mm,clip=true]{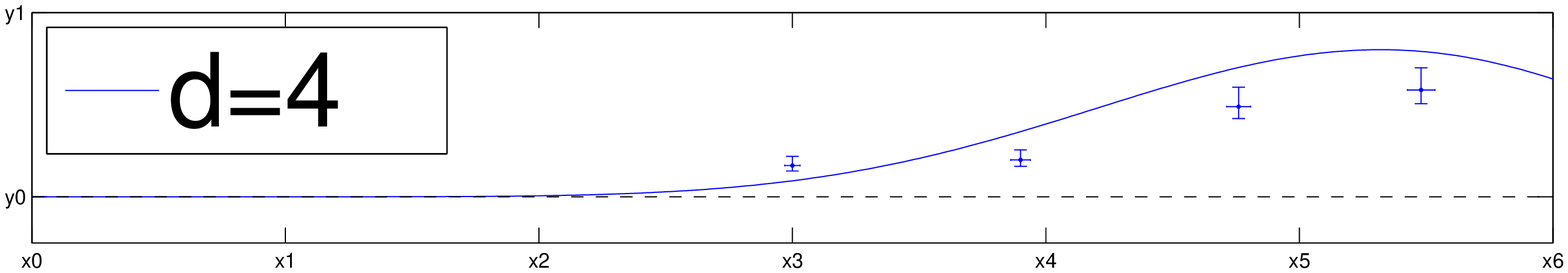}\\
\includegraphics[scale=.2,trim=0 0 0 156mm,clip=true]{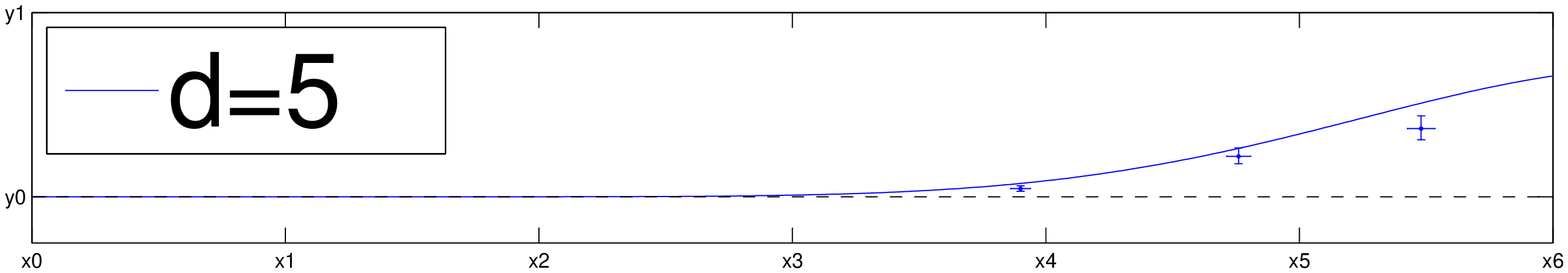}
\caption{\label{fig2}Theoretical predictions (curves) and experimental measurements (with error bars) of the normalized coincidence rate $Q(d|a,b,\Delta)\equiv Q(d|a,a,0)$ when $\Delta\approx0$ and the amplitude $a$ is scanned, for $d=0,1,2,3,4,5$. The experimental measurements are plotted entirely in terms of measured quantities, and do not depend on any adjustable parameters. Values of $a$ are deduced from measures of the RF power. Horizontal error bars are due to the limited resolution of the power meter used (we assumed a relative uncertainty on $a$ of $10^{-2}$). Vertical error bars are statistical.}
\end{figure}

\begin{figure}[ht]
\psfrag{d=0}{$d=0$}\psfrag{d=1}{$d=1$}\psfrag{d=2}{$d=2$}\psfrag{d=3}{$d=3$}\psfrag{d=4}{$d=4$}\psfrag{d=5}{$d=5$}
\psfrag{x0}[tr][cr][1][0]{\footnotesize{$\color{white}{/}\color{black}0$}}
\psfrag{x1}[tr][cr][1][0]{\footnotesize{$\color{white}{/}\color{black}\pi/2$}}
\psfrag{x2}[tr][cr][1][0]{\footnotesize{$\color{white}{/}\color{black}\pi$}}
\psfrag{x3}[tr][cr][1][0]{\footnotesize{$\color{white}{/}\color{black}3\pi/2$}}
\psfrag{x4}[tr][cr][1][0]{\footnotesize{$\color{white}{/}\color{black}2\pi$}}
\psfrag{y0}[cr][cr][1][0]{\footnotesize{0}}
\psfrag{y1}[cr][cr][1][0]{\footnotesize{0.2}}
\psfrag{y2}[cr][cr][1][0]{\footnotesize{0.4}}
\psfrag{y3}[cr][cr][1][0]{\footnotesize{0.6}}
\psfrag{y4}[cr][cr][1][0]{\footnotesize{0.8}}
\psfrag{y5}[cr][cr][1][0]{\footnotesize{1}}
\includegraphics[scale=.2,trim=0 0 0 000mm,clip=true]{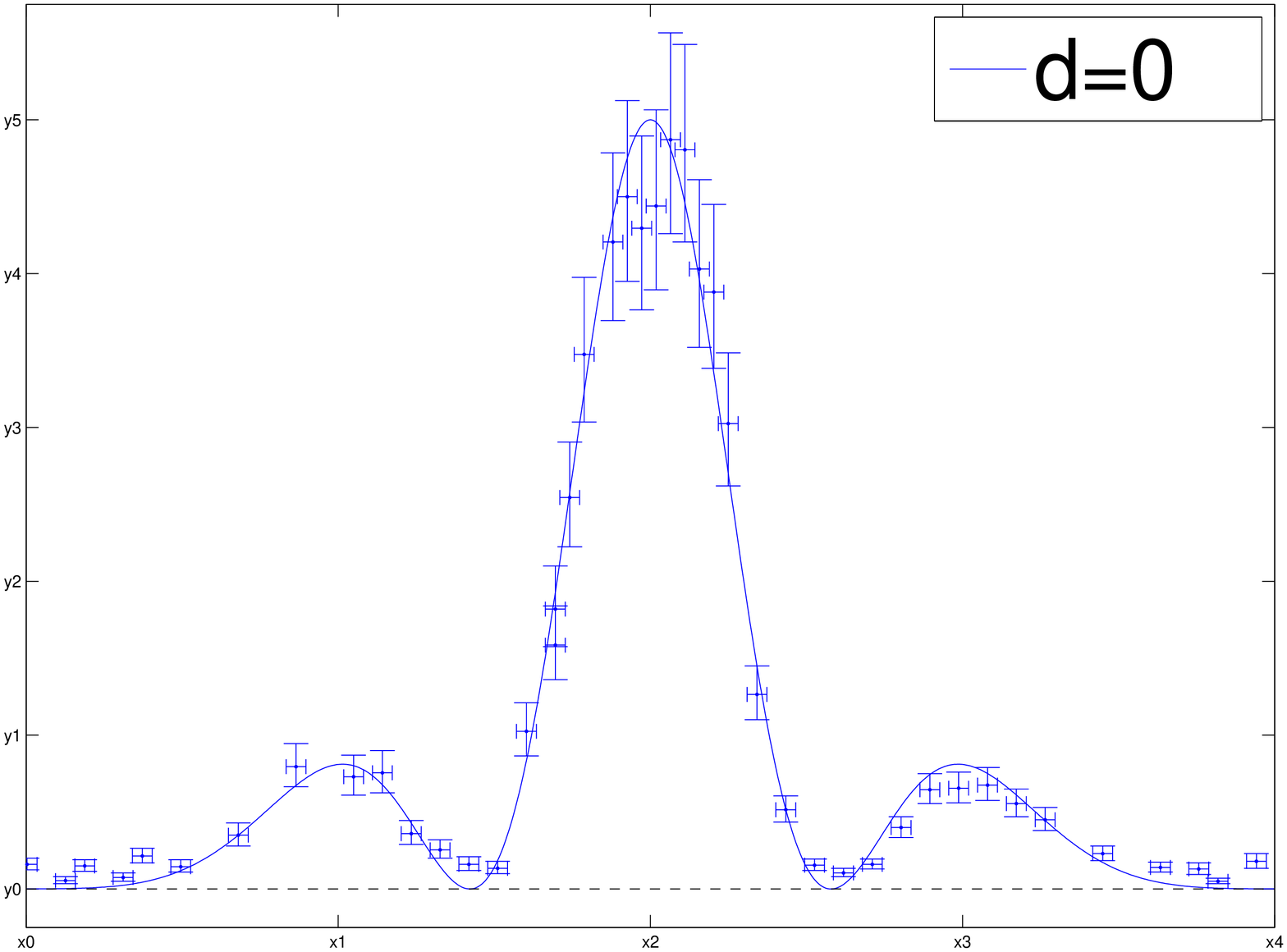}\\
\includegraphics[scale=.2,trim=0 0 0 120mm,clip=true]{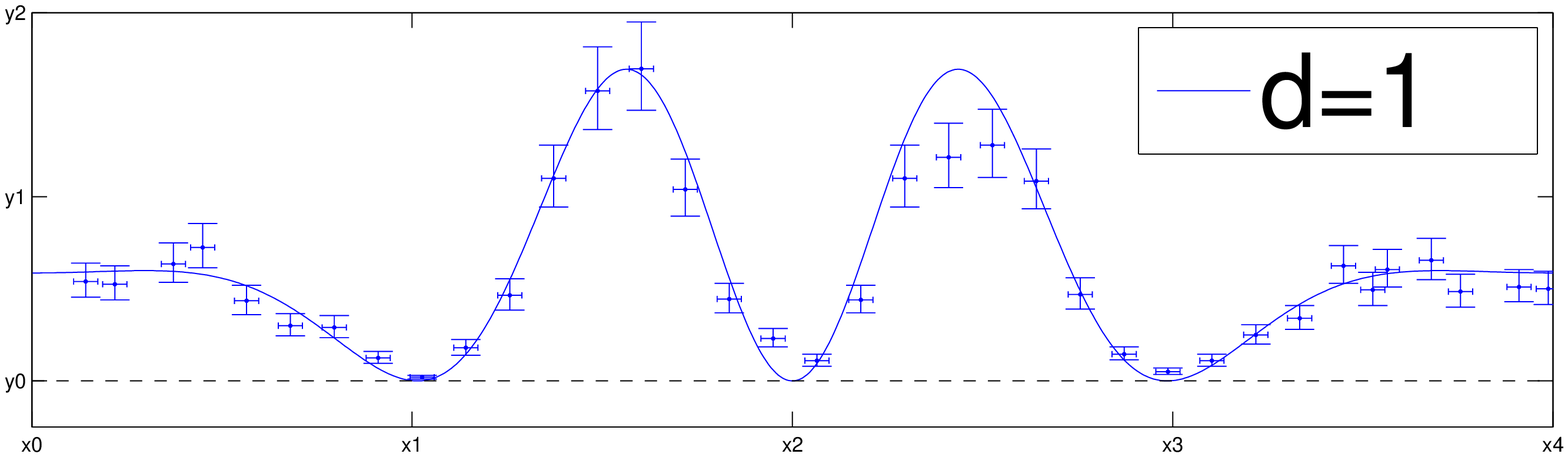}\\
\includegraphics[scale=.2,trim=0 0 0 120mm,clip=true]{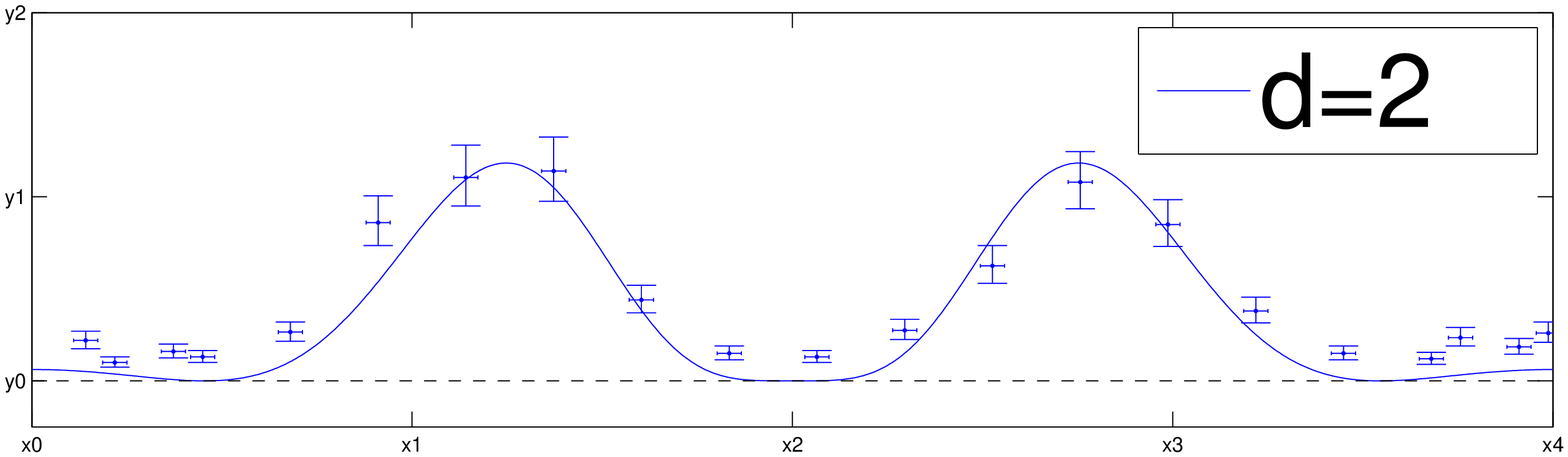}\\
\includegraphics[scale=.2,trim=0 0 0 138mm,clip=true]{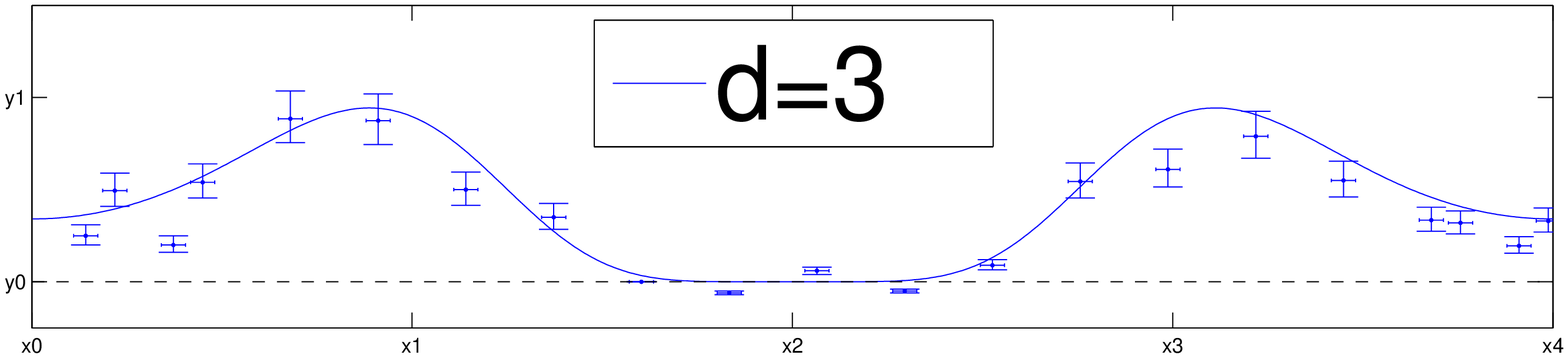}\\
\includegraphics[scale=.2,trim=0 0 0 156mm,clip=true]{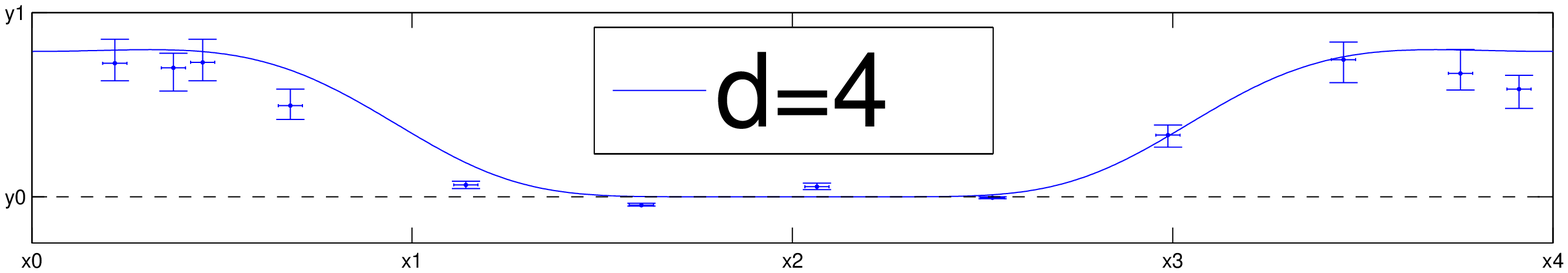}\\
\includegraphics[scale=.2,trim=0 0 0 156mm,clip=true]{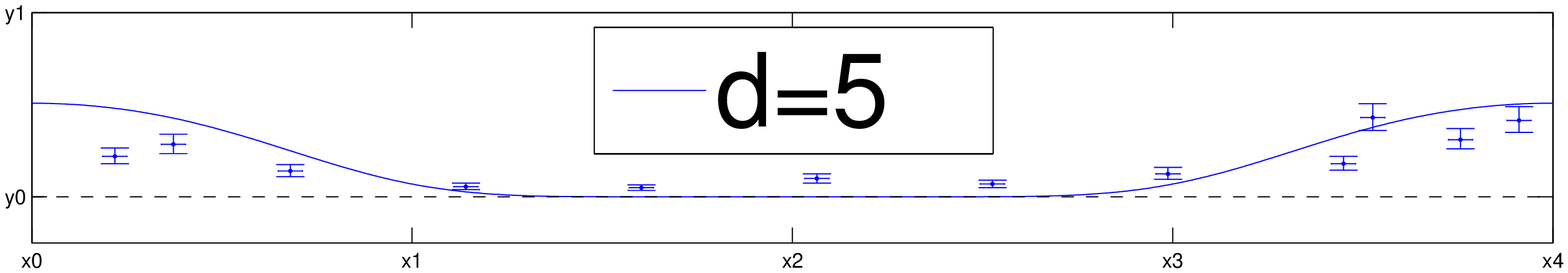}
\caption{\label{fig3}Theoretical predictions (curves) and experimental measurements (with error bars) of the normalized coincidence rate $Q(d|a,b,\Delta)\equiv Q(d|2.74,2.74,\Delta)$ when $a\approx b\approx2.74$ and the phase $\Delta$ is scanned, for $d=0,1,2,3,4,5$. To plot the experimental measurements we used the value indicated by the mechanical phase shifter -- call it $x$ --, and converted it to a phase value using the relation $\Delta=\mu x+\nu$. Parameters $\mu$ and $\nu$ were adjusted to get a good fit with the theoretical predictions. Horizontal error bars are due to the limited resolution of the phase shifter used (we assumed an absolute uncertainty on $\Delta$ of $5\cdot10^{-2}\mathrm{rad}$). Vertical error bars are statistical.}
\end{figure}

In the experiment reported below, we chose $\omega'=0$. Because the filters $\mathrm{F_{A,B}}$ have a finite bandwidth, the actual values of $\omega'$ belong to a small interval $[-\epsilon/2,\epsilon/2]$ of width approximately $3\,\mathrm{GHz}$ centered on $\omega'=0$. To resolve the frequency bins we need both that $\epsilon<\Omega$ and that the filter transmission T drops very steeply beyond $\epsilon/2$. These conditions are experimentally realized thanks to the properties of the filters, see description above and Fig. \ref{fig1} insets (a) and (b). The value of $d$ is chosen by adjusting the reflected frequency of Bob's filter to $\omega_0+d\Omega$, while the reflected frequency of Alice's filter is kept fixed on $\omega_0$. In Figs \ref{fig2} and \ref{fig3}, we compare the predictions of Eqs (\ref{eq6},\ref{eq7}) to our experimental results, for $d=0,\,1,\,2,\,3,\,4,\,5$.

Our experimental estimate, denoted $\tilde Q$, of the quantity $Q(d|a,b,\Delta)$ given by Eqs (\ref{eq6},\ref{eq7}) is obtained from the data recorded by the time to digital converter (see Fig. \ref{fig1}) by taking the total number of coincidences and subtracting the accidental coincidences, and then normalizing by the same quantity when $d=0$ and $a=b=0$:
\begin{eqnarray}\label{eq11}
\lefteqn{\tilde Q(d|a,b,\Delta)=}\nonumber\\
&&\frac{N_c(d|a,b,\Delta)-N_{ac}(d|a,b,\Delta)}{N_c(d=0|a=b=0)-N_{ac}(d=0|a=b=0)}\,.
\end{eqnarray}
This ensures that $\tilde Q$ has the same normalization as $Q$, see Eqs (\ref{eq8}, \ref{eq9}), since the coincidence rate is maximum when $a=b=0$ and $d=0$.

In Figs \ref{fig2} and \ref{fig3} the acquisition time per measured point was constant, corresponding to a number of coincidences approximately equal to $10^3P$. Experimental values are plotted with statistical vertical error bars, which is the main source of uncertainty. Note that when $d=5$ the filter $\mathrm{F_B}$ was at the limit of its tuning range, and it may have not been perfectly centered on $\omega_0+5\,\Omega$, in which case there would be a systematic underestimate of $Q(d=5|a,b,\Delta)$.

In Fig. \ref{fig2}, the normalized coincidence rate $Q(d|a,b,\Delta)$ is plotted as a function of the modulation amplitudes $a=b$ (taken to be equal) when $\Delta=0$. The number of frequency bins that interfere together is approximately given by the number of values of $d$ for which $Q$ takes a significant value, and increases when $a,b$ increase. In our experiment we were able to scan the values $a,b\in\{0,2.74\}$. When $a\approx b\approx2.74$ there are contributions from $d=0$ to $d=5$ (and by the symmetry of Eq. (\ref{eq7}) there should also be contributions from $d=-1$ to $d=-5$). This shows that at least 11 frequency bins are coherently addressed by the phase modulators.

Fig. \ref{fig3} is obtained by scanning the phase $\Delta$ when $a\approx b\approx2.74$. Note that when $\Delta=\pi$ only $d=0$ contributes, as predicted by Eq. (\ref{eq9}). The quantity $Q(d=0|a=b=2.74,\,\Delta)$ vanishes for specific values of $\Delta=\Delta^*$, see the theoretical curve in Fig. \ref{fig3}. This allows us to estimate the visibility of interferences through the usual formula $V=\left(Q_\mathrm{max}-Q_\mathrm{min}\right)/\left(Q_\mathrm{max}+Q_\mathrm{min}\right)$, where we take $Q_\mathrm{max}=P(d=0|a=b=2.74,\,\pi)$ and $Q_\mathrm{min} =Q(d=0|a=b=2.74,\,\Delta^*)$. From the data reported in Fig. \ref{fig3} we estimate that the visibility is approximately equal to $98\%$.

\section{Non Local Correlations}

Now we show how frequency bin entangled photons addressed locally by phase modulators can in principle be used to realize quantum non locality experiments. That is, we want to show that the correlations between Alice and Bob's detectors should not be explainable by a Local Hidden Variable (LHV) model.

Our starting point will be the Clauser-Horne \cite{C1974} inequality which must be satisfied by any LHV model:
\begin{eqnarray}\label{eq12}
&P(00|A_1B_1)+P(00|A_1B_2)+P(00|A_2B_1)&\nonumber\\
&-P(00|A_2B_2)\leq P(0|A_1)+P(0|B_1)\,,&
\end{eqnarray}
where $A_1,A_2$ are two possible settings of Alice's measurement apparatus and $B_1,B_2$ are two possible settings of Bob's measurement apparatus. Each measurement has two possible outcomes, and 0 denotes one of the outcomes of the measurements. The other outcome does not intervene explicitly in the inequality. Thus $P(00|AB)$ denotes the probability of both Alice and Bob finding outcomes $0$ given settings $A$ and $B$; and $P(0|A)=\sum_x P(0x|AB)$ denotes the probability of Alice finding outcome $0$. Since Alice and Bob's measurements are independent, $P(0|A)$ does not depend on $B$.

In our case the measurement settings will correspond to a choice of amplitude and phase applied to the phase modulators: $A_1=(a_1,\alpha_1)$, $A_2=(a_2,\alpha_2)$, $B_1=(b_1,\beta_1)$, $B_2=(b_2,\beta_2)$. We will take the outcome 0 in Eq. (\ref{eq12}) to correspond to the photon being registered in the frequency bin centered on $\omega_0$.

The probabilities $P(00|A_iB_i)$ can be estimated directly with our setup as they will be proportional to the number of coincidences if both filters $\mathrm{F_{A,B}}$ are centered on frequency $\omega_0$, see Eq. (\ref{eq11}).

The quantities $P(0|A_1)$ and $P(0|B_1)$ cannot be measured directly with our setup, as we only had two detectors, one on Alice's side and one on Bob's side. However we can estimate these quantities by making the following assumption (identical in spirit to the one made by Clauser-Horne in \cite{C1974}):
\begin{equation}\label{eq13}
P(0|A_1)=P(0|B_1)=P(00|a=b=0)\,.
\end{equation}
That is, we assume that the number of photons detected by Alice (Bob) in the frequency bin centered on $\omega_0$ when Alice (Bob) detector has setting $A_1$ ($B_1$) is identical to the number of coincidences in frequency bins $\omega_0$ when the phase modulators are turned off ($a=b=0$). Quantum mechanics predicts that this inequality is obeyed, since it follows from the symmetries of the correlations Eq. (\ref{eq7}), the normalization Eqs (\ref{eq8},\ref{eq9}), and the fact that the correlations are trivial when $a=b=0$, Eq. (\ref{eq10}). Our assumption is that the LHV model also obeys Eq. (\ref{eq13}).

We then insert Eq. (\ref{eq13}) into Eq. (\ref{eq12}) and divide by $P(00|a=b=0)$ to obtain
\begin{eqnarray}\label{eq14}
&\frac{P(00|A_1B_1)}{P(00|a=b=0)}+\frac{P(00|A_1B_2)}{P(00|a=b=0)}+&\nonumber\\
&\frac{P(00|A_2B_1)}{P(00|a=b=0)}-\frac{P(00|A_2B_2)}{P(00|a=b=0)}\leq 2\,.&
\end{eqnarray}
If we express that the probabilities should be proportional to the number of coincidences minus the number of accidental coincidences, we obtain the inequality
\begin{eqnarray}\label{eq15}
S&=&\tilde Q(\omega_0,\omega_0|A_1 B_1)+\tilde Q(\omega_0,\omega_0|A_1B_2)+\nonumber\\
&&\tilde Q(\omega_0,\omega_0|A_2B_1)-\tilde Q(\omega_0,\omega_0|A_2B_2)\leq2\,,
\end{eqnarray}
where $\tilde Q$ is given by Eq. (\ref{eq11}).

\begin{figure}[ht]
\psfrag{without errors}{\footnotesize{without errors}}\psfrag{with errors}{\footnotesize{with errors}}
\psfrag{x0}[tr][cr][1][0]{\footnotesize{0}}
\psfrag{x1}[tr][cr][1][0]{\footnotesize{0.5}}
\psfrag{x2}[tr][cr][1][0]{\footnotesize{1}}
\psfrag{x3}[tr][cr][1][0]{\footnotesize{1.5}}
\psfrag{x4}[tr][cr][1][0]{\footnotesize{2}}
\psfrag{y0}[br][cr][1][0]{\footnotesize{2}}
\psfrag{y1}[cr][cr][1][0]{\footnotesize{2.2}}
\psfrag{y2}[cr][cr][1][0]{\footnotesize{2.4}}
\includegraphics[scale=.2]{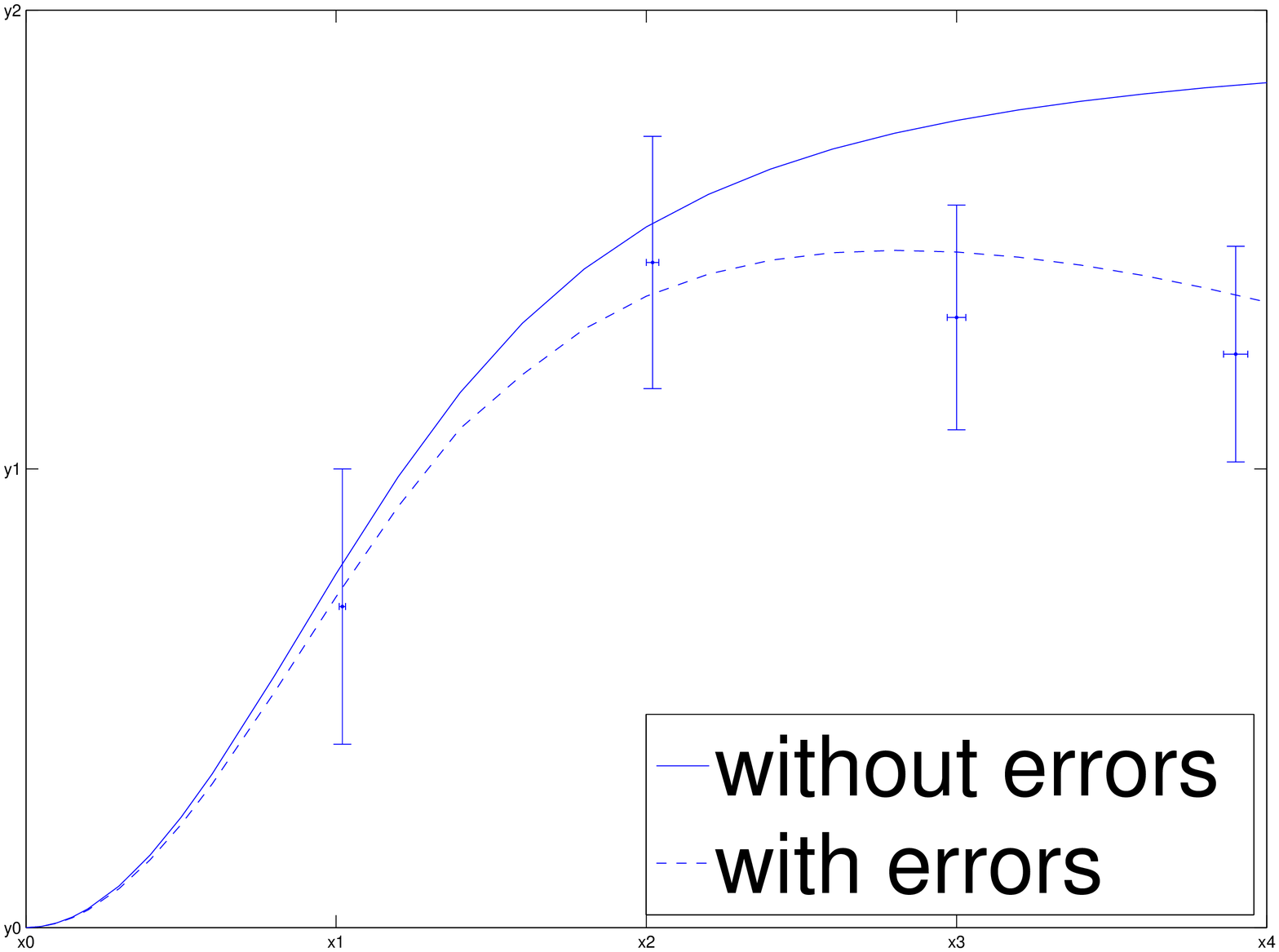}\put(-8,12){\footnotesize{$a$}}\put(-190,115){\footnotesize{$S$}}
\caption{\label{fig4}Violation of the inequality Eq. (\ref{eq15}) as a function of the modulation amplitudes $a_{1,2}=b_{1,2}=a$. The experimentally determined values of $S$ are given by the left hand side of Eq. (\ref{eq15}). They include statistical error bars (vertical axis) and RF amplitude error bars (horizontal axis). The top curve shows the theoretical evolution of the maximal value of $S$ when $\alpha_{1,2}$ and $\beta_{1,2}$ are numerically optimized. The dotted curve takes into account possible errors on $a_i,b_i,\alpha_i,\beta_i$: it shows theoretical predictions when phases and amplitudes are allowed to fluctuate around optimal values (see text for a detailed description).}
\end{figure}

On the other hand, a maximally entangled state could lead to values as high as $2\sqrt{2}$. In order to investigate whether Eq. (\ref{eq15}) can be violated experimentally, we took for simplicity the modulation amplitudes to be equal, $a_{1,2}=b_{1,2}$, and numerically optimize the phases $\alpha_{1,2},\,\beta_{1,2}$. We considered 4 different values of modulation amplitude:
\begin{itemize}
\item $a_{1,2}=b_{1,2}=0.51\,$, for which the optimal phases are $\alpha_1=0,\,\alpha_2=1.42,\,\beta_1=3.85,\,\beta_2=2.43\,$.\\
\item $a_{1,2}=b_{1,2}=1.01\,$, for which the optimal phases are $\alpha_1=0,\,\alpha_2=1.02,\,\beta_1=3.65,\,\beta_2=2.63\,$.\\
\item $a_{1,2}=b_{1,2}=1.50\,$, for which the optimal phases are $\alpha_1=0,\,\alpha_2=0.72,\,\beta_1=3.50,\,\beta_2=2.78\,$.\\
\item $a_{1,2}=b_{1,2}=1.95\,$, for which the optimal phases are $\alpha_1=0,\,\alpha_2=0.56,\,\beta_1=3.42,\,\beta_2=2.86\,$.
\end{itemize}
As one can see in Fig. \ref{fig4}, these choices lead to a strong violation of the bound 2.

The observed correlations do not however reach the theoretical optima for large values of $a=b$. The reason why it is difficult to reach the optimal value for large values of $a=b$ is that the curve $P(d=0|a=b,\,\Delta)$ is more strongly peaked around $\Delta=\pi$ for large values of $a=b$; and for the optimal values, $\Delta=\alpha-\beta$ lies on the slopes of this peak.

We have estimated what could be the effect of slight errors on $a$ and $b$ (in particular letting $b$ be slightly different of $a$), and of slight errors in the phases $\alpha_i,\beta_i$. In the experiment the error on $\beta_i$ was probably larger than the error on $\alpha_i$ because the phase shifter used to choose $\beta$ was of lesser quality. Letting $a,b$ vary by $a\cdot10^{-2}$ around the estimated value, $\alpha$ vary by $5\cdot10^{-2}$ around the ideal value, and $\beta$ vary by $10\cdot10^{-2}$ around the ideal value, and taking the worst case, we would obtain the curve indicated in dashed in Fig. \ref{fig4}. These estimates of the errors thus provide a possible explanation for the discrepancy between the theoretical optima and the observed violation of Eq. (\ref{eq15}).

\section{Conclusion}

In summary we have demonstrated how, using phase modulators and narrow band filters, one can accurately address in the frequency domain photons belonging to a high dimensional entangled state. In view of the proven success of sideband encoding for long distance QKD \cite{M9901, M9902, B2007, C2008}, this seems a promising technique for quantum communication. On the other hand, the class of unitary transformations explored in this work is somewhat limited, as it depends only on 2 parameters, see Eqs (\ref{eq1}, \ref{eq2}). However the use of non sinusoidal voltages would give rise to much more general families of unitary transformations. We hope to explore this in future work. We will also focus on studying other Bell inequalities, such as the CHSH \cite{C1969} and CGLMP \cite{C2002} inequalities.

We acknowledge support from the European Union under project QAP (contract 015848), from the Belgian Science Policy under project IAP-P6/10 (Photonics@be), from the French Agence Nationale de la Recherche under project HQNET and from the Conseil R\'egional de Franche-Comt\'e. This work also benefits from the Programme International de Coop\'eration Scientifique PICS-3742 of the French Centre National de la Recherche Scientifique.

\appendix

\section{Normalization}

We take the state $|\omega\rangle$ to be normalized to
\begin{equation}\label{a1}
\langle\omega|\omega\rangle=\delta(0)=\frac{T}{2\pi}\,.
\end{equation}
We rewrite Eq. (\ref{eq3}) as
\begin{equation}\label{a2}
|\Psi\rangle=\int_{-\infty}^{+\infty}\mathrm{d}\omega f(\omega)|\omega_0+\omega\rangle_\mathrm{A}|\omega_0-\omega\rangle_\mathrm{B}\,,
\end{equation}
where $f(\omega)$ takes into account the finite bandwidth of the signal and idler photons. We will consider the case where $f$ is a slowly varying function which can be considered approximately constant when $\omega$ changes by order $\Omega$.

Using Eq. (\ref{a1}) we have
\begin{equation}\label{a3}
\langle\Psi|\Psi\rangle=\frac{T}{2\pi}\int_{-\infty}^{+\infty}\mathrm{d}x|f(x)|^2\,,
\end{equation}
which means that the photon pairs are produced at the rate
\begin{equation}\label{a4}
R=\frac{1}{2\pi}\int_{-\infty}^{+\infty}\mathrm{d}x|f(x)|^2\,.
\end{equation}

Consider now the operator that projects onto a frequency bin of width $\epsilon$:
\begin{equation}\label{a5}
\Pi_\omega=\int_{\omega-\epsilon/2}^{\omega+\epsilon/2}\mathrm{d}\omega'|\omega'\rangle\langle\omega'|\,.
\end{equation}
The rate of coincidences in two frequency bins of width $\epsilon$ symmetrically spaced on either side of $\omega_0$ is
\begin{eqnarray}\label{a6}
\langle\Psi|\Pi_{\omega_0+\omega}^A\Pi_{\omega_0-\omega}^B|\Psi\rangle
&=&\frac{T}{2\pi}\int_{\omega-\epsilon/2}^{\omega+\epsilon/2}\mathrm{d}x|f(x)|^2\nonumber\\
&=&T\,R_{\epsilon}(\omega)\,.
\end{eqnarray}

The quantum state after it passes through the phase modulators is (see Eq. (\ref{eq4}))
\begin{eqnarray}
|\Psi'\rangle
&=&\ \int\mathrm{d}\omega\sum_{p,q}|\omega_0+\omega+p\Omega\rangle_A|\omega_0-\omega+q\Omega\rangle_B\nonumber\\
&&U_p(a,\alpha)U_q(b,\beta)f(\omega)\\
&=&\ \int\mathrm{d}\omega'\sum_{d}|\omega_0+\omega'\rangle_A|\omega_0-\omega'+d\Omega\rangle_B\nonumber\\
&&\left(\sum_pU_p(a,\alpha)U_{d-p}(b,\beta) f(\omega'-p\Omega)\right)\\
&\simeq&\ \int\mathrm{d}\omega'f(\omega')\sum_{d}|\omega_0+\omega'\rangle_A|\omega_0-\omega'+d\Omega\rangle_B\nonumber\\
&&\left(\sum_pU_p(a,\alpha)U_{d-p}(b,\beta) \right)\label{a9}\\
&=& \int\mathrm{d}\omega'f(\omega')\sum_{d}|\omega_0+\omega'\rangle_A|\omega_0-\omega'+d\Omega\rangle_B\nonumber\\
&&c_d(a,b,\alpha,\beta)\,,
\end{eqnarray}
where in obtaining line (\ref{a9}) we have used the fact that, for fixed $a,b$, $U_pU_{d-p}$ decreases rapidly with $p$ and that $f$ varies slowly so that $f(\omega'-p\Omega)\simeq f(\omega')$ for the relevant $p$. The coefficients $c_d$ are given by Eq. (\ref{eq5}).

The rate of coincidences in two frequency bins displaced one with respect to the other by $d\Omega$ is
\begin{eqnarray}
&\langle\Psi'|\Pi_{\omega_0+\omega'}^A\Pi_{\omega_0-\omega'+d\Omega}^B|\Psi'\rangle&\nonumber\\
&=|c_d(a,b,\Delta)|^2\frac{T}{2\pi}\int_{\omega-\epsilon/2}^{\omega+\epsilon/2}\mathrm{d}x|f(x)|^2&\nonumber\\
&=|c_d(a,b,\Delta)|^2\,R_\epsilon(\omega')\,T\,.&
\end{eqnarray}
Since $R_\epsilon(\omega')$ varies slowly with $\omega'$, the quantity $Q(d|a,b,\Delta)=|c_d(a,b,\Delta)|^2$ can be estimated as described in Eq. (\ref{eq11}).

Note that the normalization condition Eq. (\ref{eq8}) $\sum_d|c_d(a,b,\Delta)|^2=1$ expresses the fact that the transformation $|\Psi\rangle\to |\Psi'\rangle$ is unitary and that no photons are lost in the process.

\end{document}